\documentclass[a4paper]{IEEEtran}
\usepackage[a4paper, total={7in, 9.8in}]{geometry}
\usepackage[boxruled]{algorithm2e}
\usepackage{enumerate}
\usepackage{multirow,array}
\usepackage{cite}
\usepackage{graphicx}
\usepackage{psfrag}
\usepackage{caption}
\usepackage{subcaption}
\usepackage{url}
\usepackage{amsmath}
\usepackage{amsthm}
\usepackage{array}
\usepackage{algorithm2e}
\usepackage{amssymb}
\usepackage{amsfonts}
\usepackage{float}
\makeatletter
\renewcommand{\boxed}[1]{\text{\fboxsep=.2em\fbox{\m@th$\displaystyle#1$}}}
\makeatother

\newcommand{\B}{\boldsymbol}

\newtheorem*{conjecture*}{Conjecture}
\newtheorem{lemma}{Lemma}

\newtheorem*{corollary*}{Corollary}
\newtheorem*{remark*}{Remark}

\newtheorem{theorem}{Theorem}

\title{On Index coding for Complementary Graphs with focus on Circular Perfect Graphs}

\begin{document}
\author{
}
\date{\today}
\author{
\IEEEauthorblockN{Bhavana M, Prasad Krishnan\\}
\IEEEauthorblockA{IIIT Hyderabad, Email: \{bhavana.mvn@research., prasad.krishnan@\}iiit.ac.in\vspace{-0.5cm}}
}
\maketitle
\thispagestyle{empty}	
\pagestyle{empty}
\begin{abstract}
Circular perfect graphs are those undirected graphs such that the circular clique number is equal to the circular chromatic number for each induced subgraph. They form a strict superclass of the perfect graphs, whose index coding broadcast rates are well known. We present the broadcast rate of index coding for side-information graphs whose complements are circular perfect, along with an optimal achievable scheme. We thus enlarge the known classes of graphs for which the broadcast rate is exactly characterized. In an attempt to understand the broadcast rate of a graph given that of its complement, we obtain upper and lower bounds for the product and sum of the vector linear broadcast rates of a graph and its complement. We show that these bounds are satisfied with equality even for some perfect graphs. Curating prior results, we show that there are circular perfect but imperfect graphs which satisfy the lower bound on the product of the broadcast rate of the complementary graphs with equality. 
\end{abstract}
\section{Introduction}
\label{sec1}
 The Index Coding problem, introduced by Birk and  Kol \cite{BK} considers a system with a source and multiple receivers communicating via a noiseless broadcast channel where each receiver can have some prior knowledge about a few source symbols before the start of transmission. 
 The rate of an index code is the ratio of the length of the transmitted codeword to the length of the message symbols. An optimal index code is a transmission scheme that ensures decodability at receivers with minimal rate, known as the \textit{broadcast rate}. 
Upper and lower bounds for the  broadcast rate of index coding have been proposed through several graph-theoretic methods such as clique number \cite{BK}, chromatic number \cite{YBJ} and also via matrix methods \cite{YBJ}.  
In particular, the broadcast rate of index coding is sandwiched between the clique number $\omega(\overline{G})$ and chromatic number $\chi(\overline{G})$ of the complement of an associated graph called the \textit{side-information graph} $G$. Several approaches via  linear programming \cite{BKE}, interference-alignment  \cite{MCJ1} were also used to address this problem in new directions.

The gap between the broadcast rate of index coding and various graph theoretic parameters were investigated for instance in \cite{HN},\cite{BKE}. 
There exists a class of undirected graphs for which the clique number  equals the chromatic number for every induced subgraph $H \subseteq G $. Such graphs are termed as \textit{perfect} graphs. \emph{Imperfect graphs} are those graphs which are not perfect. It was shown in \cite{CRST} that an undirected graph $G$ is perfect if and only if $G$ contains no odd cycle of length greater than five or its complement as an induced subgraph. For a perfect graph $G$, the broadcast rate is known to be equal to $\omega(\overline{G})$. Besides perfect graphs, there are only a few classes of graphs for which the broadcast rate is known. This includes   graphs with broadcast rate $2$ \cite{BKE}
and  graphs with neighbouring side-information \cite{MCJ1}.


The contributions and organization of this paper are as follows. Our first contribution is regarding the broadcast rate of \textit{circular perfect graphs}, which were introduced by Zhu \cite{ZHU} and contains the class of perfect graphs as a strict subset. After a brief review of index coding in Section \ref{sec2}, for the case of a side-information graph $G$ which is a complement of a circular perfect graph, we show that the broadcast rate is equal to a parameter known as the \textit{circular clique number} of $\overline{G}$ (Section \ref{sec3}). In the process we also obtain a new lower bound for the broadcast rate for undirected graphs, namely the \textit{circular clique number} of $\overline{G}$. Using our result, we give the broadcast rate for several new classes of graphs in Section \ref{subsecclasses}. 

Our second contribution is regarding the relationship between the broadcast rate of a graph and its complement. Bounds on the product and sum of the chromatic numbers of a graph $G$ and its complement, which are together known as \emph{complementary graphs}, were derived by Nordhaus and Gaddum in \cite{NoG}. We present Nordhaus-Gaddum type upper and lower bounds for the product of the vector linear broadcast rates of $G$ and that of $\overline{G}$ (Section \ref{sec4}). The lower bound holds for all directed side-information graphs as well. We then show that these bounds are met with equality, even for some perfect graphs. We also curate some results from prior literature to show that there exists a class of circular perfect graphs (which include imperfect graphs) which satisfy the Nordhaus-Gaddum type lower bound with equality. We end the paper in Section \ref{discussion} with further directions of research. 

\textit{{Notations and Terminology}}: Throughout the paper the following notations are used. A graph $G$ is an undirected simple graph with vertex set $V(G)$ and edge set $E(G)$. We specifically mention that  a graph is directed if required. $K_n$ denotes the complete graph with $n$ vertices. For a positive integer $m$, $[m]$ denotes the set $\{1,2,\cdots,m\}$. The complement of graph $G$ is denoted by $\overline{G}$. A clique of $G$ is a set of vertices adjacent to each other. The \textit{clique number} $\omega(G)$ denotes the size of the largest clique of $G$. The \textit{chromatic number} $\chi(G)$ of $G$ is the minimum number of colors in a proper coloring of $G$. The \textit{fractional chromatic number} of $G$ is denoted as $\chi_f(G)$. For graphs $H$ and $G$, the notation $H\subseteq G$ denotes that $H$ is an induced subgraph of $G$. For more basic terminology related to graph theory, the reader is referred to \cite{Die}.
For a set $A,B$ $A\backslash B$ denotes the elements in $A$ but not in $B$. For some element $i$, We also denote $A\backslash \{i\}$ by $A\backslash i$. The finite field with $q$ elements is ${\mathbb F}_q$. The span of a set of vectors $U$ is denoted as $span(U)$.   

\section{Definitions}
\label{sec2}
 The single unicast index coding problem consists of a broadcast channel along with the following: 
\begin{itemize}
\item A source possessing  a set of $n$ messages $\left\{\B{x_i} : i \in [n]\right\},$ where each $\B{x_i}$ is a $t$-length vector over $\mathbb {F}_q$ denoted by $\B{x_i}=(x_{i1},\hdots,x_{it})$ where $x_{ij}\in {\mathbb F}_q$.
\item  $n$ receivers, with the receiver $i \in [n]$ demanding $\B{x_i}$.
\item The receiver $i$ knows a subset of the messages corresponding to the indices $S_i\subset[n]\backslash i$ denoted by $\{\B{x_j}: j\in S_i\}$. These are called the \textit{side information} at receiver $i$. 
\end{itemize}
The $nt$ message symbols are encoded to an $l$-length codeword $\B{c}\in{\mathbb F}_q^l$ by an encoding function and broadcasted to the receivers,  such that for each $i\in[n]$, $\B{x_i}$ can be decoded at the receiver $i$ using $\B{c}$ and $\{\B{x_j}:j\in S_i\}$. The \textit{rate} of such an \textit{index code} is then defined as $\frac{l}{t}$.

Any single unicast  problem can be defined by a directed graph $G$ called the \textit{side information graph} where $V(G)=[n]$. For $u,v\in V(G)$, there exists a directed edge in $G$ from $u$ to $v$ if $u \in S_v$. Index coding problems in which the  side information sets are symmetric, i.e. $i \in S_j$ if and only if $j \in S_i,    \forall i, j \in V$ and $i \neq j $, can be represented by undirected graphs (by replacing the directed edges between $i,j$ by undirected edges).

An index code is said to be \textit{linear} if the encoding operation is linear. If messages are of length $t=1$ the index code is called a \textit{scalar} code else the index code is known as a \textit{vector} index code. Let $\beta_q(t,G)$ denote the minimum length of any index code for a given $t$ for an index coding problem defined by a side information graph $G$. The \textit{broadcast rate} of $G$ \cite{BKE} is given as $\beta_q(G) = \lim\limits_{t\rightarrow \infty}\frac{\beta_q(t,G)}{t},$ where the limit is known to exist due to subadditivity of $\beta_q$.  In a similar fashion, let $\beta_{vl}(G)$ be the minimum rate of vector linear index codes for $G$, and $\beta_{sl}(G)$ be the minimum rate over all possible scalar linear index codes. It is known that for a given single unicast setting $G$,
$\omega(\overline{G}) \leq \beta(G) \leq \beta_{vl}(G) \leq \chi_f(\overline{G}) \leq  \chi(\overline{G})$  and $\beta_{vl}(G) \leq \beta_{sl}(G) \leq \chi(\overline{G})$ from previous studies \cite{YBJ},\cite{BKE} on bounds for $\beta(G)$.

For an index coding problem with $n$ messages, a linear index code of length $l$ encoding $t$-length messages can be represented using a matrix $B$ (over ${\mathbb F}_q$) of size $l\times nt$ such that the transmitted codeword is $B\B{x^T},$ where $\B{x}=(\B{x_1},\hdots,\B{x_n})\in {\mathbb F}_q^{nt}$ is the cumulated message vector. Assume that the columns of $B$ be denoted as $B^{ij}:i\in [n], j\in [t].$ Thus $B^{ij}$ denotes the precoding vector of the message $x_{ij}$. The following simple well known lemma (see \cite{MCJ1}, for instance, for a proof) will be used in this work to prove some results.

\begin{lemma}
\label{IAapproach}
A matrix $B$ with elements from ${\mathbb F}_q$ of size $l\times nt$ is a valid encoding matrix ensuring decoding at all receivers if and only if 

{\small
\[
B^{ij}\notin span(\{B^{ij'}:j'\in[t]\backslash j\}\cup\{B^{i'j}: i'\in[n]\backslash \{S_i \cup \{i\}\} ,  \forall j\}).
\]}
\end{lemma}

We also need the concept of a \textit{confusion graph} of an index coding problem. 
Consider the index coding  instance defined by a (possibly directed) side information graph $G$. The \textit{confusion graph} \cite{ArK1} of $G$ for $t-$length message vectors is an undirected graph ${\Gamma_t(G)}$ with vertex set ${\mathbb F}_q^{nt}$ representing the cumulated message vectors $\{(\B{x_1},\hdots,\B{x_n}):\B{x_i}\in{\mathbb F}_q^t\}$. Two vertices $\B{x}=(\B{x_1},\hdots,\B{x_n}),\B{y}=(\B{y_1},\hdots,\B{y_n})\in{\mathbb F}_q^{nt}$ are connected by an edge in $\Gamma_t(G)$ if there exists some $i\in[n]$ such that  $x_{ik} \neq y_{ik}$ while  $\B{x_j}=\B{y_j}, \forall j \in S_i $.

  \section{The Broadcast Rate of Circular Perfect Graphs}
  \label{sec3}
  In this section, we obtain the broadcast rate for the complements of a class of graphs known as circular perfect graphs, and in the process refine (albeit mildly) the $\omega(\overline{G})$ lower bound for the broadcast rate of any graph $G$. We first recall some of the basic definitions and results related to circular perfect graphs.
  
  Zhu \cite{ZHU} introduced \textit{circular perfect graphs} based on concept of a \textit{circular coloring} \cite{V}.  For positive integers $k$ and $d$,  a  $(k,d)$ \textit{circular coloring} of graph $G$ is a function
$f: V\rightarrow \{0,1,\cdots k-1\}$ with $d \leq |f(u)-f(v)|\leq (k-d)$ if $(u,v) \in E$. The \textit{circular chromatic number} $\chi_c(G) $ is defined as follows. 
\[
\chi_c(G)  =  min\left\{\frac{k}{d}: G \text{  has } (k,d) \text{ circular coloring }\right\}.
\]
A \textit{circular clique}  $K_{\frac{k}{d}} (k \geq 2d)$ is a graph $G$ with  $V(G)=\{0,\hdots,k-1\}$ and $d \leq |u-v|\leq k-d$  if and only if $(u,v)\in E(G)$. The \textit{circular clique number}  $\omega_c(G)$ is defined as 
\[
{\cal \omega}_c(G)=max\left\{\frac{k}{d} : K_{\frac{k}{d}} \subseteq G \text{ and } \gcd(k,d)=1\right\}.
\] An example of a circular clique is shown in Fig. \ref{fig:my_label}. 
\begin{figure}[h!]
    \begin{center}
    \includegraphics [width=1.2in]{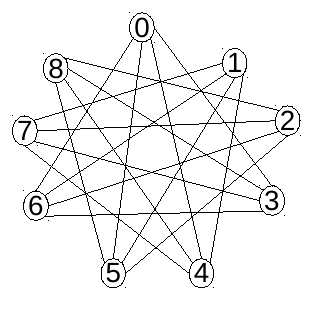}
    \label{fig:my_label}
    \end{center}
    \caption{A circular perfect graph: The circular clique $K_{9/3}$}
    \label{fig:my_label}
\end{figure}

Generally $\omega(G)\leq \omega_c(G) \leq\chi_f(G)\leq \chi_{c}(G) \leq \chi(G)$ for any  graph $G$. Furthermore it is known that $\omega(G)=\lfloor\omega_c(G)\rfloor$ and $\chi(G)=\lceil\chi_c(G)\rceil$. 

Graphs for which $ \omega_c(H)=\chi_c(H)$ for every $H \subseteq G $ are called \textit{circular perfect}. By definition, the class of perfect graphs are a subclass of the class of circular perfect graphs. However, some imperfect graphs, such as odd cycles and their complements, are also circular perfect. In Section \ref{subsecclasses}, we give many examples of graph classes which are circular perfect. For a nice survey on circular perfect graphs and some of the abovementioned results, the reader is referred to \cite{ZHU1,BJH}. 

We now prove that the broadcast rate of \textit{any} graph $G$ is bounded from below by the circular clique number $\omega_c(\overline{G})$. To show this, we need the following lemma which is known in index coding literature. Here we give a formal proof of the same as it seems missing in the literature to the best of our knowledge.
  \begin{lemma}
  \label{lowerboundconf}
  Let $H$ be a induced subgraph of a graph $G$. Then $\beta(G)\geq \beta(H)$.
  \end{lemma}
  \begin{IEEEproof}
  Consider a graph $G$ with vertex set $[n]$. Suppose $H$ be an induced sub graph with $k$ vertices say $[k]$ . We first show that for any $t$, the confusion graph $\Gamma_t(H)$ is isomorphic to a subgraph of $\Gamma_t(G).$ This would mean that $\omega(\Gamma_t(H))\leq \omega(\Gamma_t(G))$. The result will then follow from the fact shown in \cite{ArK1} (Theorem 2) where it is shown that \[
  \beta(G)=\lim_{t\rightarrow \infty}\frac{1}{t}log(\omega(\Gamma_t(G))).
  \]
  
  We now show that $\Gamma_t(H)\subseteq \Gamma_t(G)$. To do this, to each vertex $\B{v}=(\B{v_1},\hdots,\B{v_k})$ of $\Gamma_t(H)$ (where $\B{v_i}\in {\mathbb F}_q^{t}$), associate the vertex $\B{v'}=(\B{v_1},\hdots,\B{v_k},\B{z_{k+1}},\hdots,\B{z_n})\in {\mathbb F}_q^{nt}$ of $\Gamma_t(G)$ by appending some fixed $((n-k)t)$-length vector $(\B{z_{k+1}},\hdots,\B{z_n})$, where $\B{z_i} \in {\mathbb F}_q^{t}.$ Now note that if $(\B{v},\B{w})\in E(\Gamma_t(H)),$ then $(\B{v'},\B{w'})\in E(\Gamma_t(G)).$ This follows from the definition of edge set of confusion graph. Thus $\Gamma_t(H)\subseteq \Gamma_t(G)$. Hence the claim follows.

  

\end{IEEEproof}

Using Lemma \ref{lowerboundconf}, we now show the following result which gives a new lower bound of the broadcast rate of undirected side-information graphs.
\begin{lemma}
\label{lowerboundbetacirccliq}
Let $G$ be an undirected side-information graph. Then the broadcast rate of $G$ is lower bounded by the circular clique number of $\overline{G},$ i.e., $\beta(G)\geq \omega_c(\overline{G})$. 
\end{lemma}
\begin{IEEEproof}
By definition,  if $\omega_c(\overline{G})=\frac{k}{d}$, then $\overline{K_{\frac{k}{d}}}$ is an induced subgraph of $G$. The broadcast rate of $\overline{K_{\frac{k}{d}}}$  is known (Theorem VI.3, \cite{BKE}) as $\beta(\overline{K_{\frac{k}{d}}})=\frac{k}{d}.$ Using Lemma \ref{lowerboundconf}, we thus have $\beta(G)\geq \beta(\overline{K_{\frac{k}{d}}})=\frac{k}{d}=\omega_c(\overline{G}).$ This concludes the proof.
\end{IEEEproof}

Observe that for any graph $G$, $\beta(G)\leq \chi_c(\overline{G})$, since $\beta(G)\leq \chi_f(\overline{G})$ and $\chi_f(\overline{G})\leq \chi_c(\overline{G})$. Combining this observation with Lemma \ref{lowerboundbetacirccliq}, we have the following result.
\begin{theorem}
\label{circperfectthm}
Let $G$ be a side-information graph such that $\overline{G}$ is circular perfect. Then $\beta(G)=\omega_c(\overline{G})=\chi_c(\overline{G}).$
\end{theorem}
The following lemma ensures that if $G$ is the complement of a circular perfect graph, and a $(k,d)$ circular coloring of $\overline{G}$ with $\frac{k}{d}=\chi_c(\overline{G})$ is known, then the optimal code for $G$ can be obtained in polynomial time. 
 \begin{lemma}
 \label{construction}
 For any index coding problem defined by a side information graph $G$, there exists an achievable index code with rate  $\chi_c(\overline{G})$.
 \end{lemma}
 \begin{IEEEproof}
 Let  $\chi_c(\overline{G})=\frac{k}{d}$ and $C_i$ be the set of vertices colored with color $i$, for $i\in \{0,1,\cdots,k-1\}$ in a $(k,d)$ circular coloring of $\overline{G}$. We construct a linear code of rate $\frac{k}{d}$ over any field ${\mathbb F}_q$.
  Suppose that each message is a vector of length $d$. 
  Let the message demanded by receiver $v\in G$ be  $\B{x_v}=(x_{v,0}, x_{v,1},\cdots, x_{v,d-1})\in{\mathbb F}_q^d$. 
 
  For each color class $C_i$, and for each $ j \in \{0,1,\cdots, d-1\}$, define
  $X_{ij}\triangleq \sum_{v\in C_i} x_{v,j}$. Note that the receiver $v\in C_i$ which demands $x_{v,j}$ has all other messages $x_{v',j}$ for any other $v' \in C_i$ in its side information since all vertices in $C_i$ are colored with same color $i$, which means they are non-adjacent in $\overline{G}$.
  
  We now describe the transmission scheme. For each $l \in \{0,1,\cdots, k-1\},$ define by $A_l$ the set $\{k-(d-1)+l, k-(d-2)+l, \hdots,l\}(mod~k)$. For each $l$, the symbol $T_l$ defined as below is transmitted.
\[
T_l \triangleq \sum_{i\in A_l}  X_{ij_i},~where~  j_i=(k-i+l)(mod~k).
\]

We claim that the transmissions $T_l, l\in\{0,\hdots,k-1\}$ is a valid index code. To see this, note that for any $i,i' \in A_l$, $ k-(d-1) \leq |i-i'|\leq (d-1)~(mod~k)$. 
 By definition of circular coloring, the vertices $i\in A_l$ involved in transmission $T_l$ are non-adjacent to each other in $\overline{G}$ which ensures decodability of the messages in $T_l$ at the respective receivers. Also note that by definition of $T_l$, each message $x_{v,j}$ is a summand of $X_{ij}$ (such that $v\in C_i$), which is a summand in $T_{(i+j)(mod k)}$ .  Note that the rate of the code is $\frac{k}{d}$. This proves the lemma.
 
 \end{IEEEproof}

As a result of Theorem \ref{circperfectthm} and Lemma \ref{construction}, the broadcast rate and achievability of complements of circular perfect graphs are characterized. We remark that there is no known polynomial time algorithm to compute the circular chromatic number (or the circular clique number) of circular perfect graphs in general. Therefore the circular chromatic number (and the circular clique number) are known  for a few specific classes of circular perfect graphs only.  However it is known from \cite{PeW} that $\omega(G)$ and $\chi(G)$ for the circular perfect graphs can be obtained in polynomial time. As $\beta(G)=\chi_c(\overline{G})$ if $\overline{G}$ is circular perfect and $\chi(\overline{G})=\lceil\chi_c(\overline{G})\rceil$, a close approximation can be obtained for the broadcast rate of all complements of circular perfect graphs. In the following subsection, we give the circular perfect graphs for which $\chi_c(G)$ is known. 

\subsection{Classes of circular perfect graphs}
\label{subsecclasses}
We provide here some known classes of circular perfect graphs whose circular chromatic numbers are known or can be computed in polynomial time. Thus, this enlarges the set of graphs for which the broadcast rates are known. These graphs include outerplanar \cite{PW}, convex-round graphs\cite{BJH}, interlacing graphs \cite{BSBL} and few classes of concave-round graphs \cite{SC}.

\textit{Convex-round graphs} are circular perfect\cite{BJH}. A graph is convex-round if vertices of $G$ can be circularly ordered i.e $V=\{v_1,v_2,\cdots,v_n\}$ such that the neighbourhood of each  vertex $v_i$ is contiguous in the same ordering. In \cite{BJH}, it is shown that $\chi_c(G)$ for convex-round graphs can be computed in polynomial time. The class of side information graphs whose complements are convex round graphs includes the class of undirected \textit{symmetric neighbouring side information graphs}, studied for instance in \cite{MCJ1}. 

Complements of convex-round graphs are called \textit{concave-round graphs}. Some classes of circular perfect concave round graphs are shown in \cite{SC}. \textit{Webs}, which are complements of circular cliques $K_{\frac{p}{q}}$, come under the class of concave-round graphs.  For $\frac{p}{q} \geq 2$,  $\chi_c(\overline{K_{\frac{p}{q}}})= \frac{p}{\lfloor{\frac{p}{q}}\rfloor}$. It is shown in \cite{YC} that a web  $\overline{K_{\frac{p}{q}}}$ is circular perfect if and only if $q=2$ or $p=2q$ or $p=2q+1$ or $q=3,p=3k$. . In \cite{VaR}, the capacity of index coding problems with symmetric neighbouring interference is discussed. A \textit{symmetric neighbouring interference graph} is a side information graph in which there is an ordering of the $k$ vertices such that each receiver's side information set does not include $D$ messages `below' and $U$ messages `above' its desired message in the ordering. If $U=D$, then the complement of the neighbouring interference graph is a web $\overline{K_{\frac{k}{D+1}}}$ and as indicated above, we know that $\chi_c(\overline{K_{\frac{k}{D+1}}})= \frac{k}{\lfloor{\frac{k}{D+1}}\rfloor}$ for any $\frac{k}{D+1}\geq 2$. The construction for index code given in \cite{VaR}  for such neighbouring interference graphs achieves a rate of $\frac{k}{\lfloor{\frac{k}{D+1}}\rfloor}$ and is therefore optimal when the complement is circular perfect.   

Outerplanar graphs are  circular perfect and the circular chromatic number of outer planar graph is $2$ when the graph has only even cycles and is $2+\frac{1}{d}$ when $2d+1$ is length of  the shortest odd cycle\cite{PW}.
 
\textit{Interlacing graphs} are circular perfect \cite{BSBL}.
Consider a circle with points $\{1,2,\cdots,n\}$ in clockwise order. An interlacing graph is a graph with vertices $v$ corresponding to $k$- sized subsets  of $\{1, ..., n\}$ such that any two distinct circle-points in a vertex $v$ have distance at least $r$ around the circle, and an edge exists between two vertices $v,v'$ if they \textit{interlace}, i.e.  after removing the points in $v$ from the circle, the remaining points in $v'$ are in different connected components. The circular chromatic number of such an interlacing graph is $\frac{n}{k}$.

\section{On Linear Broadcast Rates for complementary Graphs}
\label{sec4}
The previous section addressed index coding problems on graphs whose complements are circular perfect. As the class of circular perfect graphs is not closed with complements, this means the broadcast rate of index coding problems for side-information graphs which themselves are circular perfect is still open. In this section, we relate the vector linear broadcast rate of a graph and its complement. Though these relationships do not seem to specialize particularly for circular perfect graphs, they seem important in the sense of providing an understanding of the broadcast rates for complementary index coding problems. These relationships can be viewed as Nordhaus-Gaddum type bounds for the vector linear broadcast rate, following the work of \cite{NoG} for $\chi(G)$.  

We first prove a lower bound on the product of vector linear rates of complementary graphs. Towards that end, we need the following lemma. The proof of this lemma may be skipped by the reader who wants to quickly move to a main result in this section, which is Theorem \ref{mainthmlowerbound}. 
\begin{lemma}
\label{lemmatensor}
Let $\B{v}$ and $\B{w}$ be non-zero vectors in a finite dimensional vector space $V$ over some field ${\mathbb F}$ such that $\B{v}\notin span(A_1)$ and $\B{w}\notin span(A_2)$, for some sets of vectors $A_1$ and $A_2$ from $V$. Let $D$ and $E$ be non-empty subsets of vectors from $V$. Then the tensor product,
\begin{align}
{\B v}\otimes{\B w}\notin span (\{\B{a}\otimes \B{d}&:\B{a}\in span(A_1), \B{d}\in D\}
\\
&\cup\{\B{e}\otimes \B{b}:\B{b}\in span(A_2), \B{e}\in E\}).
\end{align}
\end{lemma}
\begin{IEEEproof}
Without loss of generality, we assume that $D$ and $E$ are subspaces of $V$. If this is not the case, then we can use this proof to the statement of the lemma for $span(D)$ and $span(E)$ in the place of $D$ and $E$, and naturally recover our desired result. Let the dimension of $D$ and $E$ be $m$ and $n$ respectively, and their bases be $\B{d_i}:i=1,..,m$ and $\B{e_j}:j=1,...,n$ respectively. We prove the statement by induction on $n$. 

We consider the base case to be $n=0$, i.e. $E=\{\B{0}\}$. Suppose $n=0$ and the lemma is false. Then we have 
\begin{align*}
{\B v}\otimes{\B w}=\sum_{i=1}^m\B{a_i}\otimes{\B{d_i}}+\sum_{j=1}^n\B{e_j}\otimes{\B{b_j}}=\sum_{i=1}^m\B{a_i}\otimes{\B{d_i}},
    \end{align*}
    for some $\B{a_i}\in span(A_1)$ and some $\B{b_j}\in span(A_2)$. Since $\B{w}$ is a non-zero vector, from the above equation, there exists some non-zero component of the vector $\B{w}$, say the $r^{th}$ component $w_r$, such that 
    \[
    w_r\B{v}=\sum_{i=1}^m d_{ir}\B{a_i},
    \]
    where $d_{ir}$ is the $r^{th}$ component of $\B{d_i}$. This contradicts the given statement that $\B{v}\notin span(A_1).$ This proves the base case of $n=0.$
    
    Now suppose that the statement of the lemma is true for the dimension of $E$ being $n-1$. We want to prove that it is true when the dimension is $n$.
    
    Suppose not, then we have 
    \begin{equation}
        \label{eqn801}
    {\B v}\otimes{\B w}=\sum_{i=1}^m\B{a_i}\otimes{\B{d_i}}+\sum_{j=1}^n\B{e_j}\otimes{\B{b_j}}.
    \end{equation}
    As before, for some $w_r\neq 0,$ we have by the definition of the tensor product, 
$
    w_r\B{v}=\sum_{i=1}^m d_{ir}\B{a_i}+\sum_{j=1}^n b_{jr}\B{e_j}.
    $
    
    Now since $\B{v}\notin span(A_1),$ we must have some $j_1$ such that $b_{j_1r}$ is non-zero in the above equation. Thus we must have,
    \[
    \B{e_{j_1}}=l_1\B{v}+\sum_{i=1}^mg_i\B{a_i}+\sum_{j=1,j\neq j_1}^{n}h_j\B{e_j},
    \]
    for some scalars $l_1,~
    g_i:i\in\{1,...,m\},~h_j:j\in\{1,...,n\}\backslash j_1.$ Using the above in (\ref{eqn801}), we have
    \begin{align*}
    {\B v}\otimes{\B w}&=\sum_{i=1}^m\B{a_i}\otimes{\B{d_i}}+\sum_{j=1,j\neq j_1}^n\B{e_j}\otimes{\B{b_j}}+\B{e_{j_1}}\otimes\B{b_{j_1}}\\
    &=\sum_{i=1}^m\B{a_i}\otimes{\B{d_i}}+\sum_{j=1,j\neq j_1}^n\B{e_j}\otimes{\B{b_j}}+l_1\B{v}\otimes\B{b_{j_1}}\\ &~~~~+\sum_{i=1}^mg_i\B{a_i}\otimes\B{b_{j_1}}+\sum_{j=1,j\neq j_1}^{n}h_j\B{e_j}\otimes\B{b_{j_1}}.
    \end{align*}
    Thus by rearranging the terms, we have
    \begin{align}
        \nonumber
        {\B v}\otimes&({\B w}-l_1\B{b_{j_1}})\\
        \nonumber
        &\\
                \label{eqn802}   &=\sum_{i=1}^m\B{a_i}\otimes(\B{d_i}+g_i\B{b_{j_1}})+ \sum_{j=1,j\neq j_1}^n\B{e_j}\otimes(\B{b_j}+h_j\B{b_{j_1}}).
    \end{align}
        Observe with respect to (\ref{eqn802}), we have
    \begin{itemize}
        \item ${\B v}\notin span(\{\B{a_i}:i=1,...,m\})$ (by given condition).
        \item $({\B w}-l_1\B{b_{j_1}})\notin span(\{\B{b_j}+h_j\B{b_{j_1}}: j\in\{1,..,n\}\backslash j_1\})$, since $\B{w}\notin span(A_2)$ (given). 
    \end{itemize}
    Now define $E$ to be the $(n-1)$-dimensional space spanned by $\{\B{e_j}: j\in\{1,...,n\}\backslash j_1\}$, and $D$ to be the space spanned by the vectors $\{\B{d_i}+g_i\B{b_{j_1}}: i=1,...,m\}$. However, in that case, the equation (\ref{eqn802}) is a contradiction, since the induction hypothesis holds for $E$ being $(n-1)$-dimensional. Thus (\ref{eqn801}) must be false. This proves the lemma. 
\end{IEEEproof}

We now show that the product of the vector linear broadcast rates of  complementary graphs is bounded from below by the number of vertices of the graph. The analogous bound for scalar linear rates is shown in \cite{LS}.
The following lower bound holds even if the graph $G$ is directed, in which case $\overline{G}$ denotes the undirected complement. 
\begin{theorem}
\label{mainthmlowerbound}
Let $G$ be any (possibly directed) graph with $n$ vertices. Then $\beta_{vl}(G)\beta_{vl}(\overline{G})\geq n$. 
\end{theorem}
\begin{IEEEproof}
Let $\beta_{vl}(G)=\frac{l_1}{t_1}$ and $\beta_{vl}(\overline{G})=\frac{l_1}{t_2}.$ For simplicity, we only prove the statement for the case when $t_1=t_2=t$, i.e., for the case the message lengths are same in $G$ and $\overline{G}$. If not, we can assume message lengths $t_1'=t_2'=t_1t_2$, with code lengths $l_1'=l_1t_2$ and $l_2'=l_2t_1$ and continue with the proof (by time-sharing between multiple copies of the same codes, respectively for $G$ and $\overline{G}$). 

Let $B$ and $C$ be matrices of size $l_1\times nt$ and $l_2\times nt$ denoting the index coding matrices for $G$ and $\overline{G}$ respectively. In $B$, for $i\in[n], j\in [t]$, let $B^{ij}$ denote the precoding vector corresponding to the $j^{th}$ symbol of the $i^{th}$ message. Similarly we have the columns indexed in $C$. Consider the matrix $F$ formed with the columns $\{B^{ij}\otimes C^{i,j_1}: i\in[n], j,j_1\in[t]\}.$ The matrix $F$ has dimensions $l_1l_2\times nt^2$. We want to show that the rank of this matrix is $nt^2$. The theorem will then follow, since then the number of rows $l_1l_2\geq nt^2,$ and thus $\beta_{vl}(G)\beta_{vl}(\overline{G})\geq n.$ 

Let $S_i$ be the side-information of the receiver $i$ which demands the message $\B{x_i},$ in the index coding problem given by $G$. For some $i\in[n]$ and $j\in[t]$, define the sets 
\begin{align*}
A_{1,ij}=\{B^{i'j'}:&i'\in [n]\backslash \{S_i\cup\{i\}\},\forall j'\in[t]\}\\
        &\cup\{B^{ij'}: \forall j'\in[t]\backslash j\}. \\
A_{2,ij}=\{C^{i'j'}:&i'\in S_i,\forall j'\in[t]\}\cup\{C^{ij'}: \forall j'\in[t]\backslash j\}.
\end{align*}

Since $B$ and $C$ are valid index coding solutions for $G$ and $\overline{G}$ respectively, by Lemma \ref{IAapproach}, for each $i\in[n]$ and for each $j,j_1\in[t]$, it holds that  $B^{ij}\notin span(A_{1,ij})$ and $C^{ij_1}\notin span(A_{2,ij_1})$ are true. By Lemma \ref{lemmatensor}, for any non-empty sets of vectors $D$ and $E,$ we have that
\begin{align*}
B^{ij}\otimes C^{ij_1}\notin 
span&\left(\{\B{a}\otimes \B{d}:\B{a}\in span(A_{1,ij}), \B{d}\in D\}\right.
\\
&~~~\left.\cup\{\B{e}\otimes \B{b}:\B{b}\in span(A_{2,ij_1}), \B{e}\in E\}\right).
\end{align*}
Taking $D$ to be the columns of matrix $C$ and $E$ to be the columns of matrix $B$, we see that the following equation is a weaker statement of what is ensured by Lemma \ref{lemmatensor}.
\begin{multline*}
B^{ij}\otimes C^{ij_1}\\
\notin
span\Big(\{B^{i'j'}\otimes C^{i'j_1'}:i'\in[n]\backslash i,~ j',j_1'\in[t]\}\\
\cup\{B^{ij'}\otimes C^{ij_1'}:j'\in[t]\backslash j,j_1'\in[t]\backslash j_1\}
 \\
\cup\{B^{ij}\otimes C^{ij_1'}:j_1'\in[t]\backslash j_1\} \\
\cup\{B^{ij'}\otimes C^{ij_1}: j'\in[t]\backslash j\}
\Big).
\end{multline*}
Note that the above equations hold for all $i\in[n]$ and for all $j,j_1\in[t]$. This shows that each column in the matrix $F$ is independent of all the other columns. Hence the rank of $F$ is $nt^2.$ The theorem follows from the arguments in the second paragraph of the proof. \end{IEEEproof}
 \begin{remark*}
The bound in Theorem \ref{mainthmlowerbound} for $\beta_{vl}(G)\beta_{vl}(\overline{G})$ holds tight as it is clearly true if $G$ is a complete graph. Note that these graphs are perfect. However, for some circular-perfect graphs $G$ which are imperfect, the bound holds with equality. For instance, the undirected graphs with \textit{symmetric neighbouring side-information} considered in \cite{MCJ1} form complements of a subclass of convex round graphs which are circular perfect (and imperfect for certain values of $n$). Such a graph with $n$ vertices and $d-1$ symmetric side-information is shown in \cite{MCJ1} to have broadcast rate $\frac{n}{d}$. The complement of such a graph is a \textit{symmetric neighbouring interference graph} with $d-1$ symmetric non-side-information symbols. For some special cases of $n,d$, the broadcast rate of such graphs is known from \cite{VaR} as $d$. Thus these complementary graphs satisfy Theorem \ref{mainthmlowerbound} with equality.  However, in general this  bound does not hold  with equality even for perfect graphs (for instance take any bipartite graph on 3 vertices with a single edge). 
 \end{remark*}
 
 The upper bound on $\beta_{vl}(G)\beta_{vl}(\overline{G})$ is given by the following lemma.
 \begin{lemma}
  For any $G$,  $\beta_{vl}(G)\beta_{vl}(\overline{G}) \leq (\frac{n+1}{2
  })^2$ and this is tight even for a class of perfect graphs.
 \end{lemma}
 \begin{IEEEproof}
The proof of the first part follows from the fact that $\chi(G)\chi(\overline{G})\leq (\frac{n+1}{2
  })^2$ (see \cite{NoG}) and since $\beta_{vl}(G)\leq \chi(\overline{G})$. We show that there exists a class of perfect graphs for which  this bound holds with equality. Consider the perfect graph $G$ with vertices $[2n-1]$ ($n\geq 2$) obtained by taking the union of $K_n$ and $\overline{K_n}$ with one common vertex. Note that $\omega(\overline{G})=n$ and also $\chi(\overline{G})=n$.  Hence $\beta_{vl}(G) = n$. Similarly we can show that $\beta_{vl}(\overline{G})=n$. Therefore, $\beta_{vl}(G)\beta_{vl}(\overline{G})= n^2$,satisfying the bound with equality.
 \end{IEEEproof}

  \begin{lemma}
  For any $G$, $ {2}\sqrt{n} \leq \beta_{vl}(G)+\beta_{vl}(\overline{G}) \leq n+1$.\end{lemma}
 \begin{IEEEproof}
  The lower bound can be shown using Theorem \ref{mainthmlowerbound}, ($\beta_{vl}(G)\beta_{vl}(\overline{G}) \geq n)$ and the fact that $(\beta_{vl}(G)-\beta_{vl}(\overline{G}))^2 \geq 0$. Since, $\chi(G)+\chi(\overline{G}) \leq n+1$  is true by \cite{NoG} and $\beta_{vl}(G)\leq \chi(\overline{G})$ for any $G$, the upper bound holds too.
\end{IEEEproof}
\section{Discussion}
\label{discussion}
In this work we have obtained the broadcast rate of complements of circular perfect graphs. In order to tackle the question of the broadcast rate of circular perfect graphs themselves, we looked at the relationship between the broadcast rate of a graph and its complement. We obtain upper and lower bounds for the product and sum of the broadcast rates of complementary graphs. We give some example graphs for which these bounds are met with equality, but it would be worthwhile to investigate more such classes of such graphs. Another interesting line of direction is to look at broadcast rates of further generalizations of perfect graphs and their complements, beyond circular perfect graphs. 
%


\end{document}